\documentclass[aps, prl, twocolumn,superscriptaddress,nofootinbib,preprintnumbers, 10pt]{revtex4-1}
\usepackage{graphicx}
\usepackage{dcolumn}
\usepackage{amssymb,amsmath,amsthm,mathrsfs}
\usepackage{epstopdf}
\usepackage{lipsum}
\usepackage{hyperref}
\usepackage{empheq}
\usepackage{color}
\usepackage[normalem]{ulem} 
\usepackage{physics}
\usepackage{cleveref}
\usepackage[ruled,vlined]{algorithm2e}
\usepackage[normalem]{ulem}
\usepackage{comment}

\newcommand{\be}{\begin{equation}}
\newcommand{\ee}{\end{equation}}
 \newcommand{\bea}{\begin{eqnarray}}
\newcommand{\eea}{\end{eqnarray}}

\newcommand{\Mpl}{M_{{}_{\mathrm{Pl}}}}

\newcommand{\reh}{{\mathrm{reh}}}

\definecolor{burntorange}{rgb}{0.8, 0.33, 0.0}

\definecolor{blue_col}{RGB}{0,92,175}
\definecolor{red_col}{RGB}{203,64,66}
\hypersetup{linktocpage,
    colorlinks=true,
    linkcolor=red_col,
    filecolor=Mahogany,      
    urlcolor=blue_col,
    citecolor=blue_col,
}

\begin{document}

\vspace*{0.5cm}

\title{Cosmological Background Interpretation of Pulsar Timing Array Data}

\newcommand{\addressIFIC}{Instituto de F\'isica Corpuscular, Consejo Superior de Investigaciones Cient\'ificas and Universitat~de Val\`{e}ncia, 46980, Valencia, Spain}
\newcommand{\addressCERN}
{Theoretical Physics Department, CERN, 1211 Geneva 23, Switzerland}
\newcommand{\addressPisa}
{Dipartimento di Fisica ``E. Fermi'', Universit\`a  di Pisa, I-56127 Pisa, Italy}
\newcommand{\addressPd}
{Dipartimento di Fisica e Astronomia ``G. Galilei",
Universit\`a degli Studi di Padova, via Marzolo 8, I-35131 Padova, Italy}

\author{Daniel G. Figueroa} \affiliation{\addressIFIC}
\author{Mauro Pieroni} \affiliation{\addressCERN}
\author{Angelo Ricciardone} \affiliation{\addressPisa}\affiliation{\addressPd}
\author{Peera Simakachorn} \affiliation{\addressIFIC}

\date{\today}

\begin{abstract}
We discuss the interpretation of the detected signal by Pulsar Timing Array (PTA) observations as a gravitational wave background (GWB) of cosmological origin. We combine {\tt NANOGrav 15-years} and {\tt EPTA-DR2new} data sets and confront them against backgrounds from supermassive black hole binaries (SMBHBs), and cosmological signals from inflation, cosmic (super)strings, first-order phase transitions, Gaussian and non-Gaussian large scalar fluctuations, and audible axions. We find that scalar-induced, and to a lesser extent audible axion and cosmic superstring signals, provide a better fit than SMBHBs. These results depend, however, on modeling assumptions, so further data and analysis are needed to reach robust conclusions. Independently of the signal origin, the data strongly constrain the parameter space of cosmological signals, for example, setting an upper bound on primordial non-Gaussianity at PTA scales as $|f_{\tt nl}| \lesssim 2.34$ at 95\% CL. \end{abstract}


\maketitle

\preprint{CERN-TH-2023-132} 
 
%%%%%%%%%%%%%%%%%%%%%%%%%%%%%%%%%%%%%%%%%%%%%%%%%%%%
%%%%%%%%%%%%%%%%%%%%%%%%%%%%%%%%%%%%%%%%%%%%%%%%%%%%
{\bf Introduction.--} 
Pulsar timing array (PTA) collaborations, NANOGrav, EPTA/InPTA, PPTA, and CPTA,  have presented evidence~\cite{NANOGrav:2023gor, Antoniadis:2023ott, Reardon:2023gzh, Xu:2023wog} for an isotropic stochastic gravitational wave background (GWB), thanks to measuring to $\sim (3-4)\sigma$ the expected {\it Hellings-Downs} angular correlation between pulsars' line of sights~\cite{Hellings:1983fr,Jenet:2014bea}. 

This event represents a milestone in physics and marks the dawn of {\it early universe gravitational wave (GW) astronomy}, as it offers unprecedented opportunities for high-energy physics and early universe cosmology. Namely, early universe dynamics operate at energies unreachable by terrestrial means, emitting GWs that carry information about their source. These are referred to as {\it cosmological} GWBs, in contrast with {\it astrophysical} backgrounds. The detection of a cosmological GWB offers a new window into the physics {\it beyond the standard model} (BSM) that characterize the early Universe~\cite{Caprini:2018mtu}. Cosmological GWBs are expected from vacuum 
fluctuations~\cite{Grishchuk:1974ny,Starobinsky:1979ty, Rubakov:1982df,Fabbri:1983us} or particle production during inflation~\cite{Anber:2006xt,Sorbo:2011rz,Pajer:2013fsa,Adshead:2013qp,Adshead:2013nka,Maleknejad:2016qjz,Dimastrogiovanni:2016fuu,Namba:2015gja,Ferreira:2015omg,Peloso:2016gqs,Domcke:2016bkh,Caldwell:2017chz,Guzzetti:2016mkm,Bartolo:2016ami,DAmico:2021zdd,DAmico:2021vka}, and from preheating~\cite{Easther:2006gt,GarciaBellido:2007dg,GarciaBellido:2007af,Dufaux:2007pt,Dufaux:2008dn,Dufaux:2010cf,Bethke:2013aba,Bethke:2013vca,Figueroa:2017vfa,Adshead:2018doq,Adshead:2019lbr,Adshead:2019igv}, kination-domination~\cite{Giovannini:1998bp,Giovannini:1999bh,Boyle:2007zx,Li:2016mmc,Li:2021htg,Figueroa:2018twl,Figueroa:2019paj,Li:2021htg,Gouttenoire:2021wzu,Co:2021lkc,Gouttenoire:2021jhk,Oikonomou:2023qfz}, thermal plasma~\cite{Ghiglieri:2015nfa, Ghiglieri:2020mhm, Ringwald:2020ist, Ghiglieri:2022rfp}, oscillons~\cite{Zhou:2013tsa,Antusch:2016con,Antusch:2017vga,Liu:2017hua,Amin:2018xfe}, first order Phase Transitions (PhTs)~\cite{Kamionkowski:1993fg,Caprini:2007xq,Huber:2008hg,Hindmarsh:2013xza,Hindmarsh:2015qta,Caprini:2015zlo,Hindmarsh:2017gnf,Cutting:2018tjt,Cutting:2018tjt,Cutting:2019zws,Pol:2019yex,Caprini:2019egz,Cutting:2020nla,Han:2023olf,Ashoorioon:2022raz, Athron:2023mer,Li:2023yaj}, cosmic defects~\cite{Vachaspati:1984gt,Sakellariadou:1990ne,Damour:2000wa,Damour:2001bk,Damour:2004kw,Figueroa:2012kw,Hiramatsu:2013qaa,Blanco-Pillado:2017oxo,Auclair:2019wcv,Gouttenoire:2019kij,Figueroa:2020lvo,Gorghetto:2021fsn,Chang:2021afa,Yamada:2022aax,Yamada:2022imq,Kitajima:2023cek}, or large scalar fluctuations~\cite{Matarrese:1992rp,Matarrese:1993zf, Matarrese:1997ay,Nakamura:2004rm,Ananda:2006af,Baumann:2007zm,Domenech:2021ztg, Dandoy:2023jot}, see~\cite{Caprini:2018mtu} for a comprehensive review.

In this {\it Letter} we combine the NANOGrav 15-year ({\tt NG15})~\cite{NANOGrav:2023hde} and EPTA {\tt DR2new}~\cite{Antoniadis:2023utw} data sets, performing a Bayesian search for astrophysical backgrounds from supermassive black hole binaries (SMBHBs) and cosmological signals from inflation, cosmic (super)strings, first order PhTs, Gaussian and non-Gaussian scalar fluctuations, and audible axions, leading to tighter (compared to single data sets alone) constraints on the parameters of such cosmological scenarios~\cite{NANOGrav:2023hvm,Antoniadis:2023xlr}. We also provide new constraints, \emph{e.g.},~establishing an upper bound on primordial non-Gaussianity at PTA scales or stringent constraints on audible axions. We find that Gaussian and non-Gaussian scalar-induced and, to a lower degree, audible axion or superstring signals fit the data better than SMBHBs with large Bayes factors.  
This depends, however, on modeling assumptions, which highlight that more data and analysis are needed to discern between a cosmological or an astrophysical origin of the signal. 

{\bf GWB signals.--} Signal candidates to explain PTA data can be classified according to their spectrum today: 

\textit{i) SMBHBs}. The GWB spectrum at PTA frequencies ($f_{\rm yr} = 1/{\rm year}$) from SMBHBs can be parametrized as
\begin{eqnarray}\label{eq:SMBHBsignal}
\Omega_\mathrm{GW}^{(0)}(f) = 
    {\mathcal A}^{(*)}_{\rm SMBHB}\left(f/f_{\rm yr}\right)^{{2\over3}(1 + \delta)}\,,
\end{eqnarray}
with ${\mathcal A}^{(*)}_{\rm SMBHB}$ subject to large uncertainties~\cite{Sesana:2004sp,Sesana:2012ak,Burke-Spolaor:2018bvk}, $\delta = 0$ for highly populated 
circular GW-driven SMBHBs~\cite{Phinney:2001di}, and $\delta \neq 0$ in other circumstances (which include more complex shapes beyond a power law), see~\cite{NANOGrav:2023hfp} and references therein.

\textit{ii) Inflation}. Canonically-normalized  single-field slow-roll models -- {\it Vanilla} scenarios --, 
predict a 
spectrum as
\begin{equation}
\Omega_\mathrm{GW}^{(0)}(f) = \mathcal{A}^{(*)}_{\rm inf}\left({f/f_*}\right)^{n_{\rm t}}\,,    \label{eq:infl_signal}  
\end{equation} 
with a small amplitude 
$\mathcal{A}^{(*)}_{\rm inf} < 2.2\cdot10^{-16}$ at cosmic microwave background (CMB) scales around $f_* \sim 5\cdot 10^{-17}$ Hz~\cite{BICEP:2021xfz,Campeti:2022vom,Galloni:2022mok}. While vanilla models admit only a tiny red-tilt $n_{\rm t} \lesssim -0.0035$, more elaborate scenarios may develop a 'sizeable' blue tilt at small scales~\cite{Bartolo:2016ami,Guzzetti:2016mkm,Caprini:2018mtu}. 
We will use Eq.~(\ref{eq:infl_signal}) as a simple parametrization for any such scenario.

\textit{iii) PhTs}. First order PhTs can generate a GWB via bubble collisions, sound waves, and  turbulence~\cite{Hindmarsh:2013xza,Hindmarsh:2015qta,Hindmarsh:2017gnf,Weir:2017wfa, Caprini:2019egz,Cutting:2018tjt,Cutting:2019zws,Pol:2019yex,Cutting:2020nla,Cai:2023guc}. While the GWB from dark sectors can be peaked across a wide frequency range~\cite{Schwaller:2015tja}, magneto-hydrodynamic (MHD) turbulence at the QCD scale generates GWs exactly around the $\sim${\it nHz} window~\cite{Neronov:2020qrl}. Denoting $\alpha$, $\beta$, and $T_*$, the strength, (inverse) duration, and temperature of the PhT, respectively, the GWB today can be written as $\Omega_{\rm GW}^{(0)} = \Omega_{\rm rad}^{(0)}\mathcal{G}(T_*)\sum_i {\Omega}_{{\rm GW},i}$, where $\Omega_{\rm rad}^{(0)} \simeq 9.2 \cdot 10^{-5}$, $\mathcal{G}(T) \equiv ({g(T)/ g_{0}})({g_{s,0}/g_{s}(T)})^{4/3}$ weights the number of degrees of freedom at temperature $T$, $i = \rm b$ (bubbles), $i = \rm sw$ (sound waves) and $i = \rm tb$ (turbulence), 
and
\begin{eqnarray}\label{eq:spectrumPhT}
  {\Omega}_{{\rm GW},i}(f) \simeq
  \mathcal N_i \, \Delta_i(v_{\rm w}) \left({\kappa_i \, \alpha \over 1+\alpha}\right)^{p_i}
    \left({H\over\beta}\right)^{q_i}  s_i(f) \,,
\end{eqnarray}
with $\mathcal N_i$ a normalization 
factor, $p_i,q_i$ fixed exponents, $\Delta_i(v_{\rm w})$ a function of the wall velocity $v_{\rm w}$, $\kappa_i$ an efficiency factor, and $s_i(f)$ a spectral shape function, see Table~\ref{tab:PhT}.  

\textit{iv) Audible Axions (AA)}. Production of a dark-$U(1)$ gauge field~\cite{Machado:2018nqk, Machado:2019xuc, Co:2021rhi,Fonseca:2019ypl, Chatrchyan:2020pzh,Ratzinger:2020oct,Eroncel:2022vjg, Madge:2021abk} coupled via ${\alpha\over 4f_a}\phi F_{\mu\nu}\tilde{F}^{\mu\nu}$ to an oscillatory axion, can generate a large GWB~\cite{Machado:2018nqk,Guo:2023hyp}. The GWB peak frequency and amplitude depend on
axion mass $m_a$ and its decay constant $f_a$, as
\begin{eqnarray}
f_{*} \simeq 3.8 \cdot 10^{-9} \, {\rm Hz} (\alpha\theta/10)^{\frac{2}{3}} (m_a/10^{-12} \, {\rm meV})^{\frac{1}{2}}\,,\\
\mathcal{A}_* \simeq 3.41 \cdot 10^{-6} (f_a/m_p)^{4} \theta^{8/3}(10/\alpha)^{\frac{4}{3}},~~~
\end{eqnarray}
with $\alpha \gtrsim \mathcal{O}(10)$ and $\theta \sim \mathcal{O}(1)$ the initial misalignment angle \cite{Machado:2018nqk}. The spectrum today can be fitted as~\cite{Machado:2019xuc}
\begin{align}
    \Omega_{\rm GW}^{(0)}(f) = \frac{6.3 \, \mathcal{A}_*}{\left[\frac{f}{2f_{*}}\right]^{-3/2}+\exp\left[12.9\left(\frac{f}{2 f_{*}} - 1\right)\right]}.
    \label{eq:audible_axion_shape}
\end{align}

\textit{v) Cosmic Strings (CS)}. These are one-dimensional topological defects that may arise after a PhT in the early Universe~\cite{Kibble:1976sj,Vilenkin:2000jqa}, or in string theory scenarios~\cite{Witten:1985fp,Dvali:2003zj,Copeland:2003bj}. A network of CSs emits a GWB~\cite{Vilenkin:1981bx,Hogan:1984is,Vachaspati:1984gt} with spectrum today~\cite{Blanco-Pillado:2017oxo,Blanco-Pillado:2017rnf,Auclair:2019wcv}
\begin{eqnarray}
\Omega_{\rm GW}^{(0)}(f) \equiv {16\pi\Gamma (G\mu)^2\over 3pH_0^2}{1\over f}\sum_{j=1}^{\infty}{\mathcal{C}_{j}(f)\over j^{q-1}\zeta(q)}\, \,,\hspace*{0.6cm}\\ 
\mathcal{C}_{j}(f) 
= \int{dz'\over H(z')(1+z')^6}\,n\hspace*{-1mm}\left({2j\over (1+z')f},t_e(z')\right),
\label{eq:strings}
\end{eqnarray}
with $G\mu$ the string tension (in Planck units), $G$ Newton's constant, $\zeta(q)$ the {\it Riemann zeta} function, $p$ the {\it intercommutation} probability, $H(z)$ and $H_0$ the Hubble rate at {\it redshift} $z$ and today, and $n(l_e(z),t_e(z))$ the {\it number density} of loops of size $l_e(z)$ at time $t_e(z)$. To evaluate the signal we consider {\it Model I} of Ref.~\cite{Auclair:2019wcv}, with loop birth length $l = 0.1/H$~\cite{Blanco-Pillado:2013qja}, $\Gamma \simeq 50$ \cite{Blanco-Pillado:2017oxo} and $q = 4/3$ corresponding to cusps. We consider $p = 1$ for field theory strings and $10^{-3} \leq p < 1$ for superstrings.

\textit{vi) Large Scalar Fluctuations}. The coupling at second order between curvature fluctuation $\zeta$ modes leads to a GWB~\cite{Matarrese:1997ay,Nakamura:2004rm,Ananda:2006af,Baumann:2007zm,Domenech:2021ztg}. If $\zeta$ is Gaussian, the spectrum today reads 
\begingroup
\allowdisplaybreaks
\begin{eqnarray}
\Omega^{(0)}_{\rm GW}(f) = \frac{\Omega_{\rm rad}^{(0)}\mathcal{G}(\eta_c)}{24} \left(\frac{2\pi f}{a(\eta_c)H(\eta_c)}\right)^2 \overline{\mathcal{P}_h^{\rm ind}(\eta_c,2 \pi f)}\hspace*{0.2cm}\label{eq:transfer}\\
\overline{\mathcal{P}_h^{\rm ind}(\eta,k)} = 2 \int_0^\infty dt \int_{-1}^{1}ds \left[\frac{t(2+t)(s^2-1)}{(1-s+t)(1+s+t)}\right]^2 \nonumber\\
    \times~\overline{I^2(u,v,k,\eta)} \,\mathcal{P}_{\zeta}^{\rm(G)}(ku) \,\mathcal{P}_{\zeta}^{\rm(G)}(kv),\hspace*{1.34cm}
    \label{eq:guassian_induced_GW}
\end{eqnarray}
\endgroup
with $\eta$ conformal time, 
``c" indicating {\it horizon crossing}, 
and $\mathcal{P}_{\zeta}^{\rm (G)}(k)$ and $\overline{\mathcal{P}_h^{\rm ind}(\eta,k)}$ the power spectra of the curvature perturbation and of the induced tensors, with $u = (1+s+t)/2$ and $v = (1-s+t)/2$, and $\overline{I^2}$ given by Eq.~(27) of Ref.~\cite{Kohri:2018awv}. We consider two cases,
\begin{eqnarray}
 \mathcal{P}_{\zeta}^{\rm (G)}(k) = \left\lbrace
 \begin{array}{cl}
 \vspace*{2mm}
    \frac{\mathcal{A}_{ln}}{\sqrt{2 \pi \sigma_{\mathcal{R}}^2}}e^{-\frac{(\ln{k/k_*})^2}{2 \sigma_{\mathcal{R}}^2}} &  ~{\rm (Bump)}\vspace*{1mm}\\
      \mathcal{A}_{si}\Theta(k-k_{ s})\Theta(k_{l}-k) & ~ {\rm (Flat)}
 \end{array}%\nonumber
 \right.,
 \label{eq:gaussian_scalar}
\end{eqnarray}
representing a single peak ({\it Bump})~\cite{Pi:2020otn,Vaskonen:2020lbd,Kohri:2020qqd}, and a scale-invariant spectrum ({\it Flat}) within the range $k_s \leq k \leq k_l$, $k_s \ll k_l$~\cite{DeLuca:2020ioi,DeLuca:2020agl}. For broader generality, we also consider a curvature perturbation with {\it primordial non-Gaussianity} (pNG) as $\zeta({\bf x}) = \zeta_{\rm G}({\bf x}) + f_{\tt nl}(\zeta_{\rm G}^2({\bf x})- \langle\zeta_{\rm G}^2({\bf x})\rangle)$, with $\zeta_{\rm G}$ Gaussian~\cite{Cai:2018dig,Unal:2018yaa}. In this case, the GWB spectrum is given by Eqs.~(2.33)-(2.39) from Ref.~\cite{Adshead:2021hnm}.

\begin{figure*}[t]
 \includegraphics[width=0.915\linewidth]{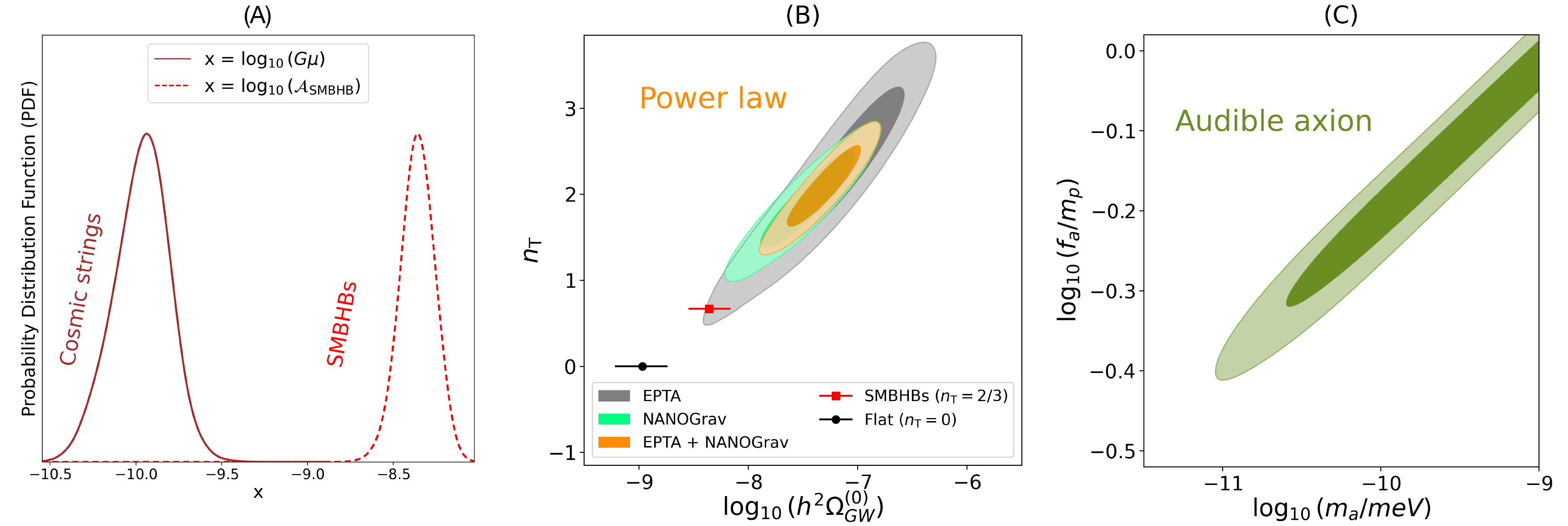}
\vspace{-1em}
\includegraphics[width=0.975\linewidth]{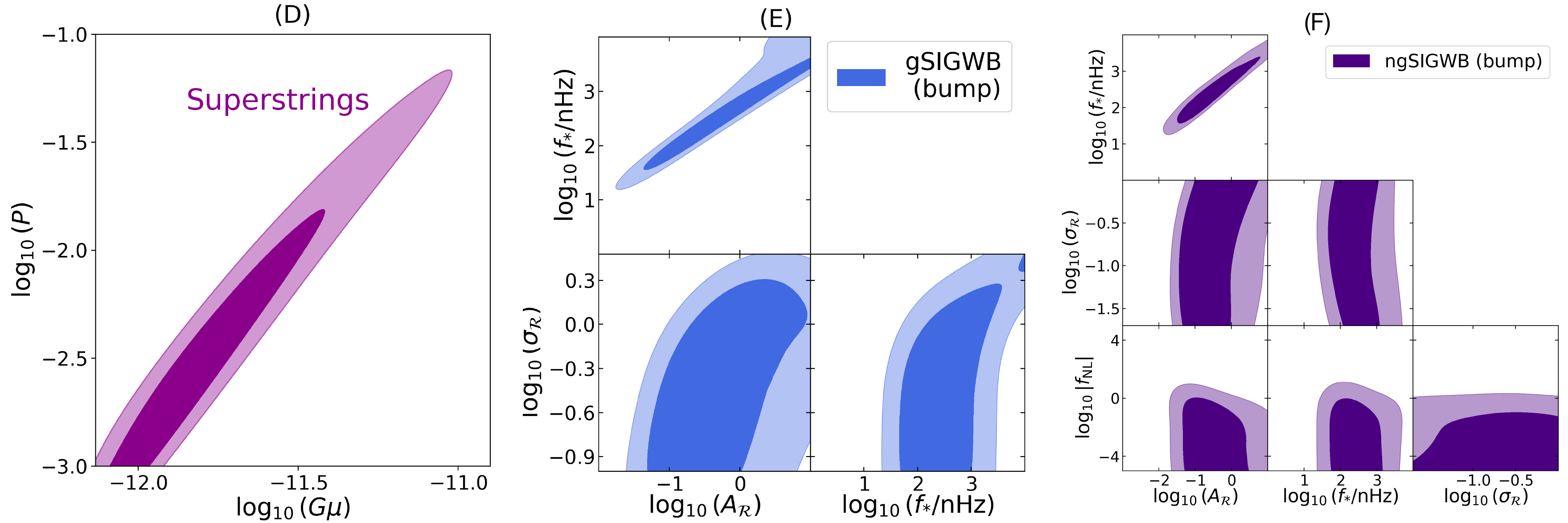}
\caption{{\it (A)} 1D posteriors of field theory CSs and SMBHBs with $n_t = 2/3$, {\it (B)} 2D posterior for a PL signal (black and gray lines indicate the error band for $n_t = 0$ and $2/3$, respectively) {\it (C)} 2D posterior for AA, {\it (D)} 2D posterior for superstrings, {\it (E)} and {\it (F)} 2D posteriors for Gaussian and non-Gaussian SIGWB bump, respectively.}
\label{fig:Constraints}
\end{figure*}

{\bf Analysis.--} For simplicity, and similarly to~\cite{InternationalPulsarTimingArray:2023mzf}, we consider the data sets {\tt NG15} and {\tt EPTA-DR2new} as independent, though we acknowledge the impact of overlapping pulsars on our findings. 
We extract physical information from free spectra chains, using Bayesian inference techniques, assuming each frequency point $f_i$ is independent, see {\it supplementary material}. In the following, we derive parameter constraints for scenarios $i)-vi)$. 

\underline{\it Power Law (PL)}. The spectrum from canonical SMBHBs, inflation, and large scalar fluctuations with {\it flat spectrum} (scenarios $(i)$, $(ii)$ and $(vi)$, respectively) correspond to $\Omega_{\rm GW}^{(0)} = \mathcal{A}_*(f/f_{\rm yr})^{n_{\rm t}}$, with free amplitude $\mathcal{A}_*$ at $f = f_{\rm yr}$ and tilt $n_t = 2/3$, free and $0$, respectively. Fitting simultaneously $\lbrace\mathcal{A}_*,n_t\rbrace$ we obtain $\mathcal{A}_* = 4.25_{-1.92}^{+3.05} \cdot 10^{-8}$ 
and $n_t = 2.08_{-0.30}^{+0.32}$, at 68\% CL. Fitting $\mathcal{A}_*$ while fixing $n_t = 2/3$ or $n_t = 0$, leads to $\mathcal{A}_*\big|_{2/3} = 4.28_{-0.97}^{+0.99} \cdot 10^{-9}$
and $\mathcal{A}_*\big|_{0} = 1.03^{+0.39}_{-0.26} \cdot 10^{-9}$, at 68\% CL. Fig.~\ref{fig:Constraints}{-(A)} shows the probability distribution function of $\mathcal{A}_{\rm SMBHB}$ (for $n_t = 2/3$). Fig.~\ref{fig:Constraints}{-(B)} shows the 1-$\sigma$ and 2-$\sigma$ contours of the full parameter space.

\underline{\it Broken Power Law (BPL)}. PhT from scenario $(iii)$ correspond to a BPL with $\Omega_{\rm GW}^{(0)}(f) \propto f^{n_1}$ and $\Omega_{\rm GW}^{(0)}(f) \propto f^{n_2}$ at low- ($f \ll f_*$) and high-frequencies ($f \gg f_*$), respectively. For each PhT contribution, the tilts $\lbrace n_1,n_2\rbrace$ are fixed. The peak's frequency $f_*$ and amplitude $\mathcal{A}_* \equiv h^2 \Omega_{\rm GW}^{(0)}(f_*)$ are set uniquely by $\alpha$, $\beta$ and $T_*$. As the data constrains directly $\lbrace f_*,\mathcal{A}_*\rbrace$, we have a clear handle on these fundamental parameters, see \emph{supplemental material}. We obtain, at 68\% CL, $T_* \in [6.2 \cdot 10^{-3},9.0]$ GeV and $\beta/H_* \in [2.2,500]$ for bubble collisions (with $\alpha \gg 1$), $T_* \in [5.5, 360]$ MeV and $\beta/H_* \in [1.0, 15]$ for a signal from sound-waves + turbulence (with $\alpha \in [0.017,0.060]$), and $T_* \in [10^{-3},3.2]$ GeV for MHD turbulence.

\underline{\it Power Law with Cutoff (PLc)}. The AA spectrum of scenario {\it (iv)} scales as $\propto f^{3/2}$ till it dies off exponentially at higher frequencies $f > f_*$. Fitting the cutoff frequency and peak amplitude $\lbrace f_*, \mathcal{A}_*\rbrace$ from Eq.~\eqref{eq:audible_axion_shape}, leads to lower bounds on the axion mass and decay constant at 68\% CL, as $m_a \gtrsim 8.0 \cdot 10^{-11}$ meV and $f_a \gtrsim 1.3 \cdot 10^{18}$ GeV, see Fig.~\ref{fig:Constraints}-(C).  

\underline{\it Cosmic Strings (CS)}. The broad spectrum in the CS scenario $(v)$, {\it cf.}~Eq.~(\ref{eq:strings}), may exhibit a different frequency dependence than a PL in the PTA window. For field theory strings we find $\log_{10}(G\mu) = -9.90_{-0.19}^{+0.11}$, at 68\% CL, see Fig.~\ref{fig:Constraints}-(A). For superstrings instead, we find $\log_{10}(G\mu) = -11.83_{-0.15}^{+0.27}$ and $\log_{10} p = -2.63_{-0.31}^{+0.49}$, at 68\% CL, see Fig.~\ref{fig:Constraints}-(D).

\begin{figure}
    \centering
    \includegraphics[width=\linewidth]{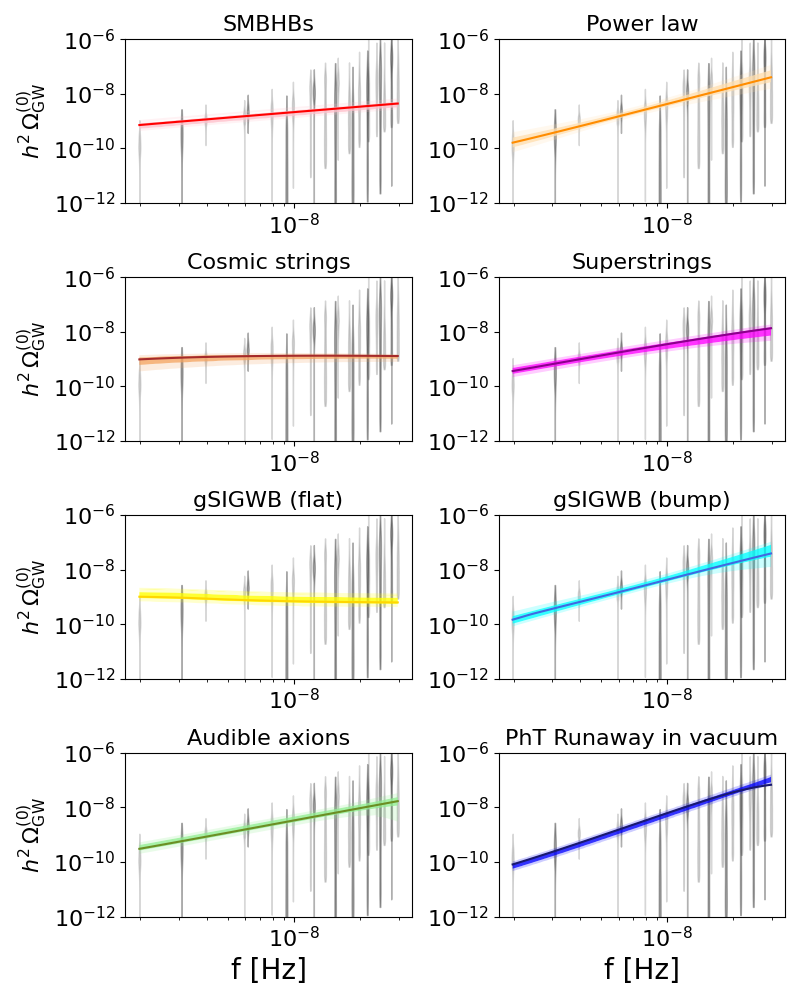}
    \caption{$h^2\Omega_{\rm GW}^{(0)}(f)$ for various cases. Solid-colored lines represent central values, while lighter (lightest) regions the 68\% (95\%) CL intervals. Dark and light gray violins are the EPTA and NANOGrav free spectrum data, respectively.}
    \label{fig:spectra_all}
\end{figure}
%%%%%%%%%%%%%%%%%%%
\underline{\it  Scalar Induced GWB (SIGWB)}. Fig.~\ref{fig:Constraints}-(E) shows the 1-$\sigma$ and 2-$\sigma$ contours, and Fig.~\ref{fig:spectra_all} the reconstructed spectrum for the best-fit parameters, with 1-$\sigma$ and 2-$\sigma$ error bands for a Gaussian bump in scenario $(vi)$. The peak frequency is constrained to be $f_* \in [0.07, 1.44] ~ {\mu \rm Hz}$, corresponding to a scale $k_{*} \in [4.57, 88]\cdot 10^{7} $ Mpc$\rm ^{-1}$, at 68\% CL. The amplitude $A_{\mathcal{R}}$ is quite large, though this depends on the value of $\sigma_{\mathcal{R}}$. We obtain $A_\mathcal{R} \in [0.10, 2.82]$ with $\sigma_\mathcal{R} \in [0.11,0.82]$ at 68\% CL.
 
\vspace*{-0.5mm}

Concerning SIGWBs with pNG, we have considered a Gaussian component with bump spectrum, considering the non-linear relation between the smoothed density contrast and the linear curvature perturbation, extended to pNG~\cite{Young:2022phe}. The parameter space is $\lbrace \sigma_\mathcal{R}$, $f_{*}$, $A_{\mathcal{R}}, f_{\tt nl}\rbrace$, subject to the perturbativity condition $A_\mathcal{R} f_{\tt nl}^2 < 1$~\cite{Adshead:2021hnm}. The data impose $|f_{\tt nl}| \lesssim 2.34$ at 95\% CL, see Fig.~\ref{fig:Constraints}-(F).

{\bf Implications and model comparison.--} We have fit PTA data to several GWB spectra. For PL inflationary signal, we can connect ${A}_*$ and $n_t$ at PTA scales to the tensor-to-scalar ratio $r$ at CMB scales, see \emph{supplemental material}. Inputting our best-fit region $\lbrace \mathcal{A}_*,n_t\rbrace$ into Eq.~\eqref{eq:GWB_inflation_in_f} leads to $r_{0.05} \in [4.37\cdot 10^{-14},3.47\cdot 10^{-9}]$ at 68\% CL (at the scale $k_* = 0.05 ~ {\rm Mpc^{-1}}$), well below current CMB constraints $r_{0.05} < 0.032$~\cite{Tristram:2021tvh}. Furthermore, a low-reheating temperature $T_{\reh} \lesssim 10.6$ GeV at 68\% CL must be necessarily imposed to respect extra radiation $\Delta N_{\rm eff}$ constraints \cite{Allen:1997ad, Smith:2006nka, Boyle:2007zx, Kuroyanagi:2014nba, Caprini:2018mtu, Ben-Dayan:2019gll}, {\it cf.}~Eq.~\eqref{eq:infl_bound_Treh}. While an inflationary signal strongly violating the consistency relation $n_t = -r/8$, with a blue tilt as large as required by PTA data, $n_t \simeq 2$, is not impossible to conceive, adding on top the requisite of a low reheating temperature, makes extremely challenging to construct a viable model. Thus, an inflationary explanation of the PTA signal looks rather implausible. 

\begin{table}
    \centering
    \begin{tabular}{r| cccc}%p{2cm}p{2cm}p{1.cm}}
    Template & \hspace{1em} ${\rm BF}_{\rm NANO}$ \hspace{0.1em} & \hspace{1em} ${\rm BF}_{\rm EPTA}$ & \hspace{0.1em} ${\rm BF}_{\rm comb}$  \hspace{0.1em} & \hspace{.1em} $\Delta{\rm AIC}_{\rm comb}$ \hspace{0.1em}\\\hline
{\bf PL($n_t = 0$)}  & 0.02 & 0.62 & 0.0014 & 10.24 \\
{\bf PL($n_t = 2/3$)}  & 2.6 & 3.1 & 1.9 & -5.20\\
{\bf PL($n_t,\mathcal{A}_*$)}  &16 & 4.1 & 180 & -12.25\\
 {\bf field th.~CS} &0.18 &  1.3 & 0.019 &5.22\\
 {\bf super CS} &29 & 3.3 & 58 &  -11.11\\
 {\bf gSIGWB} &120 & 15  & 1300 & -13.54\\
 {\bf ngSIGWB} &66 & 8.7 & 780 & -6.82\\
 {\bf AA} & 36 & 2.7 & 130 & -12.36\\
 {\bf BPL} & 17  &  1.6  & 120 & -0.19 \\
 {\bf PhTNR} & 8.5  &  8.1  & 150 & -12.85 \\
  {\bf PhTRV} & 37  &  13  & 110 & -12.34\\
  {\bf MHD} & 26  &  3.9  & 210 & -8.77 \\%\hline
    \end{tabular}
    \caption{Bayes factors $(\textrm{BF})$ for NANOGrav, EPTA, and combined data sets, and Akaike Information Criterion $(\textrm{AIC})$ for the combined data sets.
    }
    \label{tab:aic}
\end{table}

Two interesting cases emerge when fitting the data with a PL spectrum with fixed tilt $n_t$: a signal from SMBHBs with $n_{\rm t} = 2/3$~\cite{Phinney:2001di}, and a SIGWB from large Gaussian scalar fluctuations with {\it flat spectrum}, which leads to a flat spectrum in $\Omega_{\rm GW}^{(0)}$ with $n_t = 0$~\cite{DeLuca:2020ioi,DeLuca:2020agl}. We obtain that $n_t = 0$ and $n_t = 2/3$ are ruled out by the combined data sets at $5.4\sigma$ and $3.9\sigma$, respectively, see Fig.~\ref{fig:Constraints}-(B). Hence, neither of these two scenarios is preferred by the PTA data.  

We have also fit the data with a PLc spectrum $\propto f^{3/2}$ with a cutoff scale $f_*$, {\it cf.} Eq.~(\ref{eq:audible_axion_shape}). This  corresponds to an audible axion~\cite{Machado:2018nqk, Machado:2019xuc, Co:2021rhi,Fonseca:2019ypl, Chatrchyan:2020pzh, Eroncel:2022vjg, Madge:2021abk}. 
Given the lower bounds on the mass and decay constant of the axion, see Fig.~\ref{fig:Constraints}-(C), imposing a sub-Planckian $f_a \leq m_p$ motivated by string theory~\cite{Banks:2003sx}, turns PTA data into a very tight constraint as $m_a = 0.32_{-0.24}^{+1.48}$ peV.
As pointed out by~\cite{Geller:2023shn}, but contrary to previous NANOGrav 12.5-year data analysis~\cite{Madge:2023cak}, the preferred parameter region is in tension with axion over-production and $\Delta N_{\rm eff}$ constraints.

Fitting the PTA signal also admits a BPL spectrum, which can be interpreted as a GWB originated in a first-order PhT. In this case, the low- and high-frequency PL tails $\lbrace f^{n_1},f^{n_2}\rbrace$ of the spectrum depend on the dominating contribution, with $\lbrace f^{2.8}$, $f^{-1} \rbrace$ for bubbles in a strong PhT, and $\lbrace f^{3}$, $f^{-4} \rbrace$ for sound waves plus a subdominant turbulence with $\lbrace f^{2}$, $f^{-5/3} \rbrace$ in a weak PhT. We obtain a central temperature as $T_* \simeq 20$ MeV ($\simeq 94$ MeV) for a strong (weak) PhT, suggesting a first-order PhT at the MeV scale which would be in tension with lattice QCD~\cite{Aoki:2006we} and BBN prediction~\cite{Bringmann:2023opz}. In the case of magneto-hydrodynamic (MHD) turbulence, the best-fit peak position can be translated~\cite{Neronov:2020qrl} into a primordial magnetic field of strength $B_0 \simeq 23.3 \mu$G and correlation length $l_B \simeq 0.25 ~ {\rm pc}$,  which is close to current constraints on primordial magnetic fields~\cite{Planck:2015zrl, Jedamzik:2018itu, Neronov:2021xua, MAGIC:2022piy}.

Fitting the PTA signal with field theory CSs leads to a very tight constraint of the associated energy scale as $\sqrt{\mu} \simeq 1.32^{+0.20}_{-0.24} \cdot 10^{14} ~{\rm GeV}$. While this is $\mathcal{O}(10^{-2})$ smaller than standard Grand Unification scales $\sim 10^{16}~\rm GeV$, a wide variety of particle-physics models can still accommodate this value~\cite{Vilenkin:2000jqa}, see \emph{e.g.},~\cite{Kofman:1995fi,Buchmuller:2012wn, Buchmuller:2013lra,Long:2019lwl}. Besides, the GWB spectrum should be also detectable by higher frequency future experiments, enabling us to learn about the cosmological history and particle physics above the BBN $\sim$MeV scale~\cite{Cui:2017ufi, Cui:2018rwi, Gouttenoire:2019kij, Gouttenoire:2019rtn, Gouttenoire:2021wzu, Gouttenoire:2021jhk, Auclair:2019jip, Auclair:2021jud, Kitajima:2022lre, Ghoshal:2023sfa}. As discussed later, the GWB from field theory CS does not fit well the PTA data. A cosmic superstring network, on the other hand, can fit very well the PTA data, thanks to accommodating a smaller energy scale $\sqrt{\mu} \simeq 1.56^{+0.48}_{-0.39} \cdot 10^{13} ~{\rm GeV}$, while suppressing the intercommutation probability to $p = [1.13,7.06] \cdot 10^{-3}$. 

Finally, we have fit PTA data with SIGWBs choosing $\sigma_{\mathcal{R}} <3$ as required by CMB constraints \cite{Inomata:2018epa}. The data indicate that a GWB created by a Gaussian curvature perturbation with a bump spectrum (gSIGWB-bump) requires a peak frequency $f_* \in [0.07, 1.44] ~ {\mu \rm Hz}$ at 68\% CL. We compute the primordial black hole (PBH) abundance considering the non-linear relation between curvature and matter perturbations~\cite{Young:2019yug}, and find a PBH distribution peaked around masses $M_H/M_{\odot} \in [1.5\cdot 10^{-5},5.6\cdot 10^{-3}]$, 
see \emph{supplemental material}. Fig.~\ref{fig:SIGW_constraint} shows how the region of PTA best-fit is cut by the PBH over-abundance criterion, as well as by gravitational lensing, and LIGO/VIRGO bounds~\cite{Green:2020jor, bradley_j_kavanagh_2019_3538999}. As the observational constraints are stronger than the over-abundance bound, our analysis implies that the PTA signal cannot correspond to an interpretation of PBH as the dark-matter totality. In any case, the best-fit PTA region can only survive for $\sigma_\mathcal{R}> 1$. Future PBH searches could clarify whether the SIGWB could explain the origin of the PTA signal. We have also applied an analogous analysis for the case of a large curvature fluctuation with local pNG and Gaussian component with a bump spectrum (ngSIGWB-bump), see Fig.~\ref{fig:Constraints}-(F). Assuming perturbativity as  $A_{\mathcal{R}}f_{\tt nl}^2 < 1$, PTA data provides a stringent constrain on the pNG parameter as $|f_{\tt nl}| \lesssim 2.34$ at 95\% CL. This is a remarkable constraint on pNG at PTA scales, which we highlight are $\sim 8$ orders of magnitude smaller than CMB scales~\cite{Planck:2019kim}. Our bound is conservative compared to bounds from PBH abundance, which depend strongly on the PBH formation and evolution details.

To determine the most favored models we have performed a model selection/comparison Bayesian analysis.  We employed two different criteria: the Bayes Factor (BF) $B_{i} = \rm{evidence}[\mathcal{H}_i]/\rm{evidence}[\mathcal{H}_{SMBHB}]$ (where $\mathcal{H}_i$ represents each model $i$ and SMBHB a reference model based on the 2D Gaussian prior from~\cite{NANOGrav:2023hvm})
and the Akaike Information Criterion given by $\textrm{AIC} \equiv - 2 \ln \mathcal{L} + 2 k$ (where $k$ is the number of parameters of the template) and compared to SMBHB from~\cite{NANOGrav:2023hvm}.  We report the values for all models analyzed in Table~\ref{tab:aic}. We obtain that both for NANOGrav and EPTA, gSIGWB-bump shows the largest BFs, ${\rm BF}_{\rm NANO}=120$ and ${\rm BF}_{\rm EPTA}=15$, followed up closely by ngSIGWB-bump, with ${\rm BF}_{\rm NANO}=66$ and ${\rm BF}_{\rm EPTA}=8.7$. Additionally, the BF and $\Delta$AIC of the combined data indicate a very good fit for both gSIGWB and ngSIGWB signals. There is also ``substantial'' evidence for the AA and cosmic superstring models, both with ${\rm BF}_{\rm NANO} \sim 30$ and ${\rm BF}_{\rm EPTA} \sim 3$.
While the superstring signal fits the data well, as pointed out  in~\cite{NANOGrav:2023hvm,Antoniadis:2023xlr}, the AA cannot explain the combined data because of its own relic constraints. We also remark that, as already found in~\cite{NANOGrav:2023hvm}, a simple PL with $n_t = 2/3$ exhibits small BFs of order $\sim 2-3$ (also for the combined data set), indicating a poor/low quality fit. The remaining models examined in this paper show lower BF and AIC values, suggesting a reduced ability to fit the new data.

From our analysis, we can conclude that some cosmological models fit the data better than SMBHBs (assuming the simplest model for the latter, see below). Among these, the gSIGWB+bump provides the best fit, with very strong evidence in NANOGrav data and strong in EPTA data, compared to SMBHBs. Decisive evidence shows up when the data sets are combined, with BF$_{\rm comb} \simeq 1300$, though this number should not be taken too literally, given the caveats of our analysis (\emph{e.g.} overlapping pulsars). Similarly, SIGWB with pNG also fits quite well the data, with BF$_{\rm comb} \simeq 780$, and most importantly, leading to a stringent bound at the PTA scales as $ |f_{\tt nl}| \lesssim 2.34$ at 95\% CL. Concerning other cosmological models, we observe that, in agreement with Refs.~\cite{NANOGrav:2023hvm,Antoniadis:2023xlr}, field theory strings do not fit well the data, whereas superstrings exhibit a good fit (though not as good as SIGWB signals). However, as the superstring signal scaling $\propto 1/p$ is not well established, we advise taking this case with care. If eccentricity, environmental, or population effects are considered, the GWB spectrum is no longer a simple power law, and it might become similar to a broken power law, leading to an improvement of the SMBHBs fit~\cite{NANOGrav:2023hfp,Antoniadis:2023xlr,Ellis:2023dgf}, although such modifications are model dependent and currently under discussion. Therefore, no clear conclusion about the origin of the signal can be reached at this point. While the detection of a GWB by PTA collaborations~\cite{NANOGrav:2023gor, Antoniadis:2023ott, Reardon:2023gzh, Xu:2023wog} has opened an exciting window for astrophysics, early universe cosmology, and BSM physics, further data and analysis are still needed to reach robust conclusions. 

{\it Note added}: Taken individually, some of our results overlap with the new physics searches by NANOGrav and EPTA collaborations~\cite{NANOGrav:2023hvm,Antoniadis:2023xlr}, as well as
with recent studies~\cite{Franciolini:2023pbf,Ellis:2023tsl,Franciolini:2023wjm,Vagnozzi:2023lwo,Cai:2023dls}. There are however some differences. For example, ours is a NANOGrav and EPTA combined data Bayesian analysis, which improves some of the model constraints. Furthermore, we have considered a smoothing procedure of the GWB spectra per bin.
Model-wise, we have extended the search to audible axions and SIGWBs with pNG. The day before we submitted our work, two studies appeared in the ArXiv considering PTA constraints on local pNG~\cite{Liu:2023ymk,Wang:2023ost}, both presenting similar constraints on $f_{\tt nl}$.

{\it Acknowledgements --} {We thank Stas Babak, Matthias Koschnitzke, Andrea Mitridate, Gabriele Perna, Hippolyte Quelquejay Leclere, Kai Schmitz, G{\'e}raldine Servant, and Zach Weiner for useful discussions. DGF (ORCID 0000-0002-4005-8915) is supported by a Ram\'on y Cajal contract with Ref.~RYC-2017-23493. This work was supported by Generalitat Valenciana grant PROMETEO/2021/083, and by  Spanish Ministerio de Ciencia e Innovacion grant PID2020-113644GB-I00. AR acknowledges financial support from the Supporting TAlent in ReSearch@University of Padova (STARS@UNIPD) for the project “Constraining Cosmology and Astrophysics with Gravitational Waves, Cosmic Microwave Background and Large-Scale Structure cross-correlations’'.}

\bibliography{auto,other}

\clearpage
%\appendix
\onecolumngrid

\renewcommand{\thepage}{S\arabic{page}}
\renewcommand{\theequation}{S.\arabic{equation}}
\renewcommand{\thetable}{S.\Roman{table}}
\renewcommand{\thefigure}{S.\arabic{figure}}
\setcounter{page}{1}
\setcounter{equation}{0}
\setcounter{table}{0}
\setcounter{figure}{0}

%%%%%%%%%%%%%%%%%%%%%%%%%%%%%%%%%%%%%%%%%%%%%%%%%%%%%%%%%%%%%%%%%%%%%%%%%%%%%%%%%%%%%%%%%%%%%%%%%%%%

\begin{center}
\textit{\Large Supplemental Material
}
\end{center}
\noindent

This supplemental material contains more details on the data analysis used in the main text. The translation from the GWB templates' parameters into the particle physics and cosmological parameters is also provided. As examples, we link the power-law and broken power-law templates to the GWB from primordial inflation and first-order PhT, respectively.
Lastly, we discuss briefly the primordial black hole and the scalar-perturbation power-spectrum constraints relating to the scalar-induced GWB interpretation.

 {\bf Assumptions.--} In all our computations, we have assumed standard $\Lambda$CDM cosmology for the cosmic history, with $\Omega_{\rm rad}^{(0)} \simeq 9.2 \cdot 10^{-5}$, $h \simeq 0.68$, and the evolution of the relativistic degrees of freedom $g_*(T)$ and $g_{*,s}(T)$ obtained from App.C of \cite{Saikawa:2018rcs}. The (reduced) Planck mass is $m_p \equiv (8\pi G)^{-1/2} \simeq 2.44\cdot 10^{18}$ GeV. 

{\bf Input Data.--} The effect of GWB is characterized, in the PTA time-of-arrival data, as the time-correlated stochastic process, parametrized by the one-sided power spectral density $S(f_i = i/T_{\rm obs})$ for each frequency bin that is a harmonic $i^{\rm th}$ of the inverse of observational time $T_{\rm obs}^{-1}$ \cite{NANOGrav:2023hvm}. This turns into the power spectrum of the fraction energy density in GW today as \cite{Allen:1997ad},
\begin{align}
    \Omega_{\rm GW}^{(0)}(f) = \frac{8 \pi^4 f_{\rm yr}^5}{H_0^2} S(f)\left(\frac{f}{f_{\rm yr}}\right)^{5},
\end{align}
In the case in which the signal PSD can be parametrized with a power law like, we have
\begin{eqnarray}
    S(f) = \frac{A_{\rm CP}^{2}}{12\pi^2}\left(\frac{f}{f_{\rm yr}}\right)^{-\gamma} {f_{\rm yr}}^{-3}~~~~~ \Leftrightarrow ~~~~ \Omega_{\rm GW}^{(0)}(f) = \frac{2 \pi^2 f_{\rm yr}^{2}}{3H_0^2}A_{\rm CP}^{2} \left(f\over f_{\rm yr}\right)^{5-\gamma} \,.
\end{eqnarray}
The ``violins" in the plots represent the free spectrum marginalized probability distribution function (PDF) of the amplitude of the sine-cosine Fourier pair (i.e., $\sqrt{S(f)/T}$ in seconds) for all frequencies $f_i$'s.

\section{More details on data analysis}
This section provides a more detailed explanation of the data analysis technique employed in this work. For what concerns the likelihood definition, we proceed as follows. We start by performing a Gaussian kernel smoothing of the histograms built from the free spectrum chains.
The result of this procedure is used to estimate probability distribution functions (PDFs) for signal amplitude in the frequency bins $f_k$, which are thus assumed to be independent from one another. In particular, under these assumptions, the full log-likelihood is represented as 
\begin{equation}
    \label{eq:log_likelihood}
    \ln \mathcal{L}=\sum_{k} \ln p_{k}(\mathcal{D}_{k}| \vec{\theta},) \; ,
\end{equation}
where $\vec{\theta}$ is the parameter vector of the models, and $\mathcal{D}_{k}$ and $p_{k}$ are the data chains and likelihoods respectively. Some signals, such as the scalar-induced GWB, exhibit variations on frequency scales that are (much) smaller than the experiment resolution $\Delta f = 1/T_{\rm tot}$ (with $T_{\rm tot}$ as the total observation time). Since detection is based on integrated power within frequency bins of width $\Delta f$, the data cannot capture these variations. To account for this, we smooth the templates by averaging over bins of size $\Delta f$. The log-posterior is defined as the sum of the log-likelihood in~\cref{eq:log_likelihood} and of the (log)-priors, defined in the following section. Finally, the parameter space is sampled using the Nested Sampler \texttt{Polychord}~\cite{Handley:2015fda, Handley:2015vkr} via the \texttt{Cobaya}~\cite{Torrado:2020dgo} interface, as well as the MCMC sampler~\cite{Lewis:2013hha}. 

{\bf Prior.--} For each model, we assign uniform/log-uniform priors to all the parameters, ensuring wide ranges to avoid impact on parameter estimation. A collection of the ranges used in the analysis is reported in~\cref{tab:priors}.

{\bf White-noise.--} In analogy with the approach of~\cite{NANOGrav:2023hvm,Antoniadis:2023ott}, in the full frequency analysis, we assume that a flat (white noise-like) component might be present in the data. In the referenced works, this additional component was included on top of a low-frequency power law behavior, giving rise to a broken power law model with a flat tilt at large frequencies. In the present work, we add this secondary component to all the models we discuss. This procedure automatically selects the relevant bins to constrain the cosmological signal from the full frequency range and models the high-frequency part of the spectrum in terms of the extra flat component. We have compared the compatibility of the results obtained with the full frequency analysis (including the additional flat component) with the results obtained using only the low frequency part of the spectrum and found good agreement for most models.

\section{From Templates to Particle-Physics and Cosmological Parameters}

This section provides the dictionary between the templates' parameters and the information about particle physics and cosmology. In the main texts, some of the templates are already parametrized as the particle physics parameters, \emph{i.e.,} CSs $G\mu$, the audible axion $\{m_a, f_a\}$, and the scalar-induced GW with the parameters of the curvature-perturbation power spectrum.
So we focus on the power-law template, which explains GWB from primordial inflation, and the broken power law, which is compatible with first-order phase transitions in many scenarios.
The central value and the error intervals of the physical parameters, quoted in the main text, are obtained by doing an analysis directly on such parameters.

{\bf Power-law.--} Apart from the astrophysical GWB from SMBHBs whose spectral tilt is fixed to $2/3$, the GWB from primordial inflation beyond the vanilla slow-roll inflation can lead to an arbitrary tilt.

\begin{figure}[ht!]
  \begin{minipage}[c]{0.47\textwidth}
    \includegraphics[width=\textwidth]{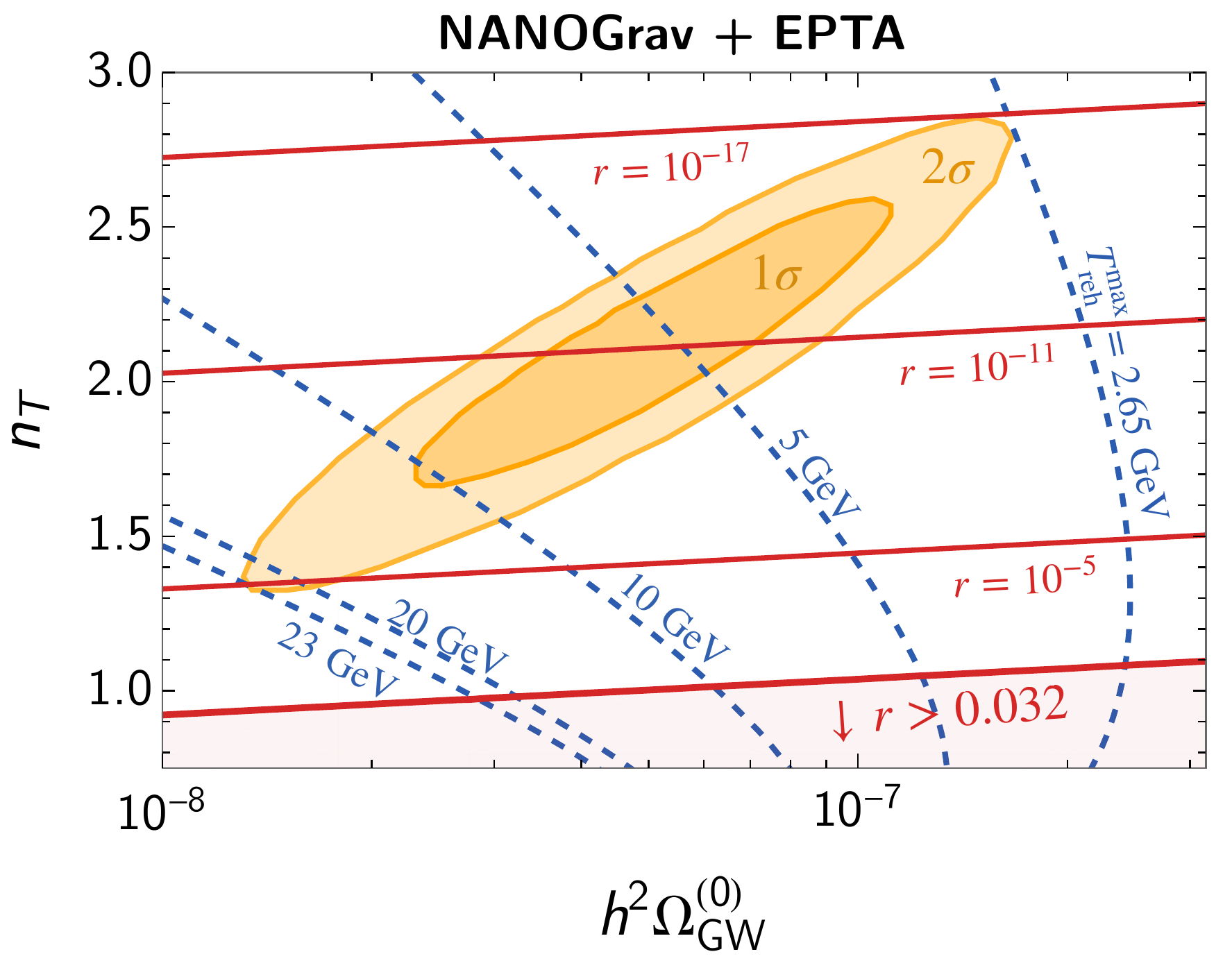} 
  \end{minipage}\hfill
  \begin{minipage}[c]{0.5\textwidth}
	\caption{The CMB observations tightly constrain the PTA-signal interpretation as the inflationary GWB (blue contours). The bound on the tensor-to-scalar ratio  (in red) excludes the region with $r > 0.032$. The BBN-$N_{\rm eff}$ bound  requires a low-scale reheating: $T_{\rm reh} < 10.6 $ GeV, at 68\% CL. 
    }
 \label{fig:inf_cmb_cons}
  \end{minipage}
\end{figure}

\textit{Primordial Inflation.}
The spectrum of GW energy density fraction today (assuming the $\Lambda$CDM cosmology) reads \cite{Caprini:2018mtu},
\begin{align}
    h^2 \Omega_{\rm GW}^{(0)}(k) =  \frac{3}{128} \Omega_{\rm rad}^{(0)} h^2 \mathcal{P}_t(k) \left[\frac{1}{2} \left(\frac{k_{\rm eq}}{k}\right)^2 + \frac{16}{9}\right] \xrightarrow{k \gg k_{\rm eq}} \frac{\Omega_{\rm rad}^{(0)} h^2}{24} \mathcal{P}_t(k),
    \label{eq:GWB_inflation_in_k}
\end{align}
with $\mathcal{P}_t(k)$ the primordial tensor power spectrum, and $k_{\rm eq} \simeq 1.3 \cdot 10^{-2} ~ {\rm Mpc^{-1}}$ the comoving scale at matter-radiation equality. The second step is justified because the comoving scale around the PTA scale is much larger than $k_{\rm eq}$, \emph{i.e.,} the GW wavenumber-frequency conversion is $k = 2 \pi f$ or,
\begin{align}
    k = 6.25 \cdot 10^{5} ~ {\rm Mpc}^{-1} \left(\frac{f}{\rm nHz}\right).
    \label{eq:comoving_scale_freq}
\end{align}
With a power-law primordial tensor perturbation, the GW spectrum is also a power-law. We parametrize it in terms of the CMB measurement at the pivot scale $k_{\rm pivot} = 0.05 ~{\rm Mpc^{-1}}$,
\begin{align}
    \mathcal{P}_t(k) = r A_s \left(\frac{k}{k_{\rm pivot}}\right)^{n_t},
    \label{eq:tensor_power_spec_param}
\end{align}
where $r$ is the tensor-to-scalar ratio whose upper bound is $r \leq 0.032$ {\small (Planck, BICEP/Keck, BAO)} \cite{Tristram:2021tvh}, $A_s$ is the primordial scalar perturbation measured at $k_{\rm pivot}$ whose value is constrained to $A_s \simeq 2.1 \cdot 10^{-9}$ {\small (TT, TE, EE+low-E+lensing)} \cite{Planck:2018jri}, and $n_t$ is the spectral tilt as in Eq.~\eqref{eq:infl_signal}.
We can rewrite Eq.~\eqref{eq:GWB_inflation_in_k} in terms of frequency as, using Eqs.~\eqref{eq:comoving_scale_freq} and \eqref{eq:tensor_power_spec_param},
\begin{align}
    h^2 \Omega_{\rm GW}^{(0)}(f) \simeq 1.19 \cdot 10^{-16} \left(\frac{r}{0.032}\right) \left(\frac{A_s}{2.1 \cdot 10^{-9}}\right) \left(\frac{f}{f_{\rm pivot}}\right)^{n_t},
    \label{eq:GWB_inflation_in_f}
\end{align}
where $f_{\rm pivot} \simeq 8 \cdot 10^{-17}$ Hz is the frequency corresponding to the pivot scale $k_{\rm pivot}$.
For given amplitude $ \Omega_{\rm GW}^{(0)}(f_*)$ and frequency $f_*$, an increasing $n_t$ gives a smaller $r$.
From the analysis in the main text, the best-fitted spectrum is highly red-tilted and corresponds to the tensor-to-scalar ratio of $r\simeq 10^{12}$, which is firmly excluded by the CMB bound.  
Nonetheless, the 2$\sigma$ region prefers the blue-tilted spectrum and has the tensor-to-scalar ratio of $r \in [5 \times 10^{-4}, 0.032]$; see Fig.~\ref{fig:inf_cmb_cons}.
The highly blue-tilted (large $n_t$) GWB could be in tension with the extra-radiation bound. The total GW energy density fraction should not exceed that of the extra radiation \cite{Allen:1997ad, Smith:2006nka, Boyle:2007zx, Kuroyanagi:2014nba, Caprini:2018mtu, Ben-Dayan:2019gll} parametrized by the number of extra neutrino species $\Delta N_\nu$,
\begin{align}
    \int_{f_{\rm BBN}}^{f_{\rm reh}}\frac{df}{f} \Omega_{\rm GW}^{(0)} \leq 5.6 \cdot 10^{-6} \Delta N_\nu ~ \xrightarrow{\rm Eq.~\eqref{eq:GWB_inflation_in_f}} f_{\rm reh} \lesssim f_{\rm BBN} \left[1 + (4.3 \cdot 10^{9}) \, n_t \left(\frac{f_{\rm pivot}}{f_{\rm BBN}}\right)^{n_t} \left(\frac{\Delta N_\nu}{0.2}\right) \left(\frac{0.032}{r}\right) \left(\frac{2.1 \cdot 10^{-9}}{A_s}\right) \right]^{\frac{1}{n_t}},
    \label{eq:bbn_bound_inflation}
\end{align}
where we integrated from the reheating scale $f_{\rm reh}$ to BBN (or CMB) scale $f_{\rm BBN}$ in the second step.
Using that the perturbation re-enters the horizon at $k= a(T) H(T)$ when the Universe has temperature $T$, we obtain the conversion,
\begin{align}
    f = 1.84 ~ {\rm nHz} \left(\frac{T}{100 ~ \rm MeV}\right)\left(\frac{g_*(T)}{10.75}\right)^{\frac{1}{2}}\left(\frac{g_{*,s}(T)}{10.75}\right)^{-\frac{1}{3}}.
    \label{eq:freq_temp_relation}
\end{align}
The BBN scale $T_{\rm BBN}\simeq \rm MeV$ corresponds to $f_{\rm BBN} \simeq 1.84 \cdot 10^{-11}$ Hz.
From Eq.\eqref{eq:bbn_bound_inflation}, the extra-radiation bound becomes a constraint on the reheating temperature of the Universe,
\begin{align}
    T_{\rm reh} \lesssim ~  &{0.68 ~ \rm MeV} \left(\frac{f_{\rm BBN}}{1.84 \cdot 10^{-11} ~ \rm Hz}\right) \left(\frac{106.75}{g_*(T_{\rm reh})}\right)^{\frac{1}{2}}\left(\frac{106.75}{g_{*,s}(T_{\rm reh})}\right)^{-\frac{1}{3}} \times \nonumber\\
    & \times \left[1 + (4.3 \cdot 10^{9}) \, n_t \left(\frac{f_{\rm pivot}}{f_{\rm BBN}}\right)^{n_t} \left(\frac{\Delta N_\nu}{0.2}\right) \left(\frac{0.032}{r}\right) \left(\frac{2.1 \cdot 10^{-9}}{A_s}\right) \right]^{\frac{1}{n_t}}.
    \label{eq:infl_bound_Treh}
\end{align}
For a power-law GW spectrum in PTA range -- $h^2 \Omega_{\rm GW}^{(0)}(f_*) \simeq 10^{-10}$ and $f_* \simeq \rm nHz$ -- with $A_s \simeq 2.1\cdot 10^{-9}$ and $r = 0.032$, the spectral tilt is $n_t \simeq 0.73$ and the bound on the reheating temperature reads $T_{\rm reh} \lesssim 44 ~ \rm TeV$. 
From Fig.~\ref{fig:inf_cmb_cons}, the tip of the 2$\sigma$ region that survives the CMB constraint on $r$ requires a low-scale reheating temperature $T_{\rm reh} < 0.1 - 1$ TeV.

Lastly, the required $T_{\rm reh}$ in Eq.~\eqref{eq:infl_bound_Treh} does not guarantee that the extra-radiation bound is not further violated at $f> f_{\rm reh}$.
Because the completion of reheating\footnote{When the Universe starts its radiation-dominated phase.} and the end of inflation can be separated by the \emph{reheating phase}, which is not necessarily radiation-dominated. Assuming the dominant energy density is a fluid of equation-of-state $\omega$, the GWB from inflation is affected through its transfer function and receives a tilt \cite{Giovannini:1998bp, Giovannini:1999bh, Boyle:2007zx},
\begin{align}
    \Omega_{\rm GW}^{(0)}(f) \propto \mathcal{P}_{t}(f) f^{-2(1-3\omega)/(1+3\omega)} \xrightarrow{\rm Eq.~\eqref{eq:tensor_power_spec_param}} \Omega_{\rm GW}^{(0)}(f) \propto f^{n_t -2(1-3\omega)/(1+3\omega)}.
    \label{eq:eos_bound}
\end{align}
If the extra-radiation bound is saturated at $T_{\rm reh}$, the reheating phase is required to have $\omega \lesssim (2 - n_t)/(6 + 3 n_t)$. 

{\bf Broken power law.--} GWBs from short-lasting sources, \emph{e.g.}, first-order PhT, and the particle production of the audible axion, usually have a peaked shape. 
Since the UV tail of the audible axion's spectrum is not a power law but an exponential cut-off in Eq.~\eqref{eq:audible_axion_shape}, the broken power-law template represents only the case of PhT\footnote{Although the generic spectrum can appear as many peaks because of the mixing among three contributions, we will only focus on the simplified scenarios.}.
The peak position is composed of the peak amplitude today,  $\Omega_{\rm GW}^{(0)} = \Omega_{\rm rad}^{(0)}\mathcal{G}(T_*)\sum_i {\Omega}_{{\rm GW},i}$ with $\mathcal{G} \equiv [g_*(T_*)/g_*(T_0)][g_{*,s}(T_*)/g_{*,s}(T_0)]^{-4/3}$ and ${\Omega}_{{\rm GW},i}$ in Eq.~\eqref{eq:spectrumPhT}, and the peak frequency today $f_*$ defined in Tab.~\ref{tab:PhT}.
We can thus convert the fitted template's parameters into the PhT parameters in the three scenarios: a) the bubble-dominated case for a strong PhT, b) the sound-wave-dominated case for a weak PhT, and c) the turbulence-dominated for a medium strength.
Moreover, we also consider case c) the turbulence from magneto-hydrodynamic (MHD) eddies, whose GW spectrum relates to the primordial magnetic field and which might or might not rely on PhT.

\begin{table*}[h!]
\begin{tabular}{cccc}
\hline\\[-0.75em]
 & ~ ~  Bubbles \cite{Huber:2008hg} ~ ~ & ~ ~ Sound-wave \cite{Hindmarsh:2015qta} ~ ~ & ~ ~ Turbulence \cite{Caprini:2009yp} ~ ~ \\[0.25em] \hline\\[-0.5em]
$\mathcal{N}_i$ & 1 & 0.16 & 20 \\[0.5em]
$\Delta_i(v_{\rm w})$ & $\frac{0.11 \, v_{\rm w}^3}{0.42 + v_{\rm w}^2}$ & $v_{\rm w}$ & $v_{\rm w}$ \\[0.5em]
$p_i$ & 2 & 2 & 3/2 \\[0.5em]
$q_i$ & 2 & 1 & 1 \\[0.5em]
$s_i(f, f_*)$ & $\frac{3.8 (f/f_*)^{2.8}}{1 + 2.8 (f/f_*)^{3.8}}$ & $\left(\frac{f}{f_*}\right)^{3} \left[\frac{7}{4 + 3 (f/f_*)^{2}}\right]^{\frac{7}{2}}$ & $\frac{(f/f_*)^3}{(1 + f/f_*)^{11/3}\left(1 + 8 \pi f/\tilde{H}_*\right)}$ \\[0.5em]
$f_{*}$ & $11 {\rm nHz}  \left(\frac{0.62}{1.8 - 0.1 v_{\rm w} + v_{\rm w}^2}\right) \mathcal{F}$ & $13  {\rm nHz} ~ v_{\rm w}^{-1} \mathcal{F}$ & $19  {\rm nHz} ~  v_{\rm w}^{-1} \mathcal{F}$ \\ [0.5em] \hline
\end{tabular}
\caption{Parameters and functions---defined around Eq.~\eqref{eq:spectrumPhT} in the main text---for three types of the GWB from first-order PhTs, taken or derived from Ref.~\cite{Breitbach:2018ddu}.}
\label{tab:PhT}
\begin{align}
    {\rm with} ~ ~ \mathcal{F} &\equiv \left(\frac{\beta}{H_*}\right) \left(\frac{T_*}{0.1 ~ {\rm GeV}}\right) \left(\frac{g_*(T_*)}{10.75}\right)^{\frac{1}{2}} \left(\frac{g_{*,s}(T_*)}{10.75}\right)^{-\frac{1}{3}},\\
    {\rm and} ~ ~ \tilde{H}_* &\equiv H_* \left(\frac{a_*}{a_0}\right) \simeq 11 ~ {\rm nHz} \left(\frac{T_*}{0.1 ~ {\rm GeV}}\right) \left(\frac{g_*(T_*)}{10.75}\right)^{\frac{1}{2}} \left(\frac{g_{*,s}(T_*)}{10.75}\right)^{-\frac{1}{3}}.
\end{align}
\end{table*}

\textit{a) Bubbles.}  According to the \emph{envelope approximation} \cite{Huber:2008hg}\footnote{Nonetheless, many numerical simulations and other models found slightly different values ranging from $f^{[-3, -0.9]}$ and $f^{[-3, -1]}$ for the IR and UV slopes, respectively. See Fig.~6.3.1 of Ref.~\cite{Gouttenoire:2022gwi} for the summary and references therein.}, the shape of the GWB has the IR tail with $\Omega_{\rm GW} \propto f^{2.8}$ and the UV tail with $\Omega_{\rm GW} \propto f^{-1}$. If there is a vacuum-dominated epoch or the so-called \emph{super-cooling phase} ($\alpha \gg 1$), a strong first-order PhT happens in a vacuum, and the collisions among bubble walls source the dominant GW signal.
With Eq.~\eqref{eq:spectrumPhT} and Tab.~\ref{tab:PhT}, we translate the peak position -- the peak frequency today $f_*$ and the amplitude $\Omega_{\rm GW}^{(0)}(f_*)$ -- of the GWB into the PhT parameters:
\begin{align}
    \left(\frac{\beta}{H_*}\right)_{\rm b} &\simeq 206 \left[\Delta_{\rm b}(v_{\rm w}) \kappa_{\rm b}^2 \mathcal{G}(T_*) \left(\frac{10^{-9}}{h^2 \Omega_{\rm GW}^{(0)}(f_*)}\right)\right]^{1/2},\\
    \left(T_*\right)_{\rm b} &\simeq 0.48 ~ {\rm MeV} \left(\frac{f_*}{10 ~\rm nHZ}\right) \left(\frac{h^2 \Omega_{\rm GW}^{(0)}(f_*)}{10^{-9}}\right)^{1 \over 2} \left(\frac{g_{*,s}(T_*)}{g_*(T_*)}\right) \left(\frac{1.8 - 0.1 v_{\rm w} + v_{\rm w}^2}{0.62}\right) \Delta_{\rm b}^{- \frac{1}{2}} \kappa_{\rm b}^{-1}.
\end{align}
Typically, the vacuum energy is efficiently converted into the energy of the bubbles $\kappa_{\rm b} = 1$, and the wall velocity reaches the speed of light $v_{\rm w} \to 1$ (and $\Delta_{\rm b} \approx 0.08$). 
For the range compatible with PTA observation, this strong PhT following the vacuum-dominated phase completes at $T_* \sim$MeV and can obstruct BBN \cite{Bringmann:2023opz}. 

\textit{b) Sound wave.} According to Ref.~\cite{Hindmarsh:2015qta}\footnote{The recent simulations also found the presence of the intermediate scale -- corresponding to the width of the sound shell --  with $\propto f^{1}$ behavior near the peak.}, the shape of the GWB has the IR tail with $\Omega_{\rm GW} \propto f^{3}$ and the UV tail with $\Omega_{\rm GW} \propto f^{-4}$. The sound-wave contribution dominates when the PhT is not too strong, and the bubbles expand in the background thermal plasma and reach the terminal velocity (so-called \emph{non-runaway} scenario). The peak position in Eq.~\eqref{eq:spectrumPhT} and Tab.~\ref{tab:PhT} can be translated into PhT parameters ($\alpha$, $\beta/H_*$, $T_*$) by,
\begin{align}
    \left(\frac{\beta}{H_*}\right)_{\rm sw} &\simeq 68  \left(\frac{10^{-9}}{h^2 \Omega_{\rm GW}^{(0)}(f_*)}\right) \mathcal{G}(T_*) v_{\rm w} \left[\frac{(1+\alpha)^2\alpha^{-2} \kappa^{-2}_{\rm sw}(\alpha)}{100}\right]^{-1},\\
    \left(T_*\right)_{\rm sw} &\simeq 1.5 ~ {\rm MeV} \left(\frac{f_*}{10 ~\rm nHZ}\right) \left(\frac{h^2 \Omega_{\rm GW}^{(0)}(f_*)}{10^{-9}}\right) \left[\frac{10.75}{g_*(T_*)}\right]^{\frac{3}{2}} \left[\frac{g_{*,s}(T_*)}{10.75}\right]^{\frac{5}{3}} \left[\frac{(1+\alpha)\alpha^{-1} \kappa^{-1}_{\rm sw}(\alpha)}{10}\right]^2.
\end{align}
where the efficiency factor is $\kappa_{\rm sw}(\alpha) \simeq \alpha(0.73 + 0.083 \sqrt{\alpha} + \alpha)^{-1}$ \cite{Espinosa:2010hh}. The numerator of the last bracket in the $T_*$ expression is $\simeq 84$ and 6.2 for $\alpha = 0.1$ and 0.5, respectively.
Note that there is degeneration among $\alpha$, $\beta/H_*$, and $T_*$, which could be further broken by considering a UV completion of this low-$T$ first-order PhT. 

\textit{c) Turbulence.} From  Ref.~\cite{Caprini:2009yp}, the shape of the GWB is a double-broken power-law spectrum. The slope far from the peak in the IR is $\Omega_{\rm GW} \propto f^{3}$ from causality. The intermediate IR scale $\tilde{H}_* < f <f_*$ with $\Omega_{\rm GW} \propto f^{2}$ corresponding to the time scale of the turbulence, and the UV tail with $\Omega_{\rm GW} \propto f^{-5/3}$. 
Typically during PhT, the GW from turbulence is accompanied by the more dominant sound-wave contribution (i.e., $\kappa_{\rm turb} < \kappa_{\rm sw}$). In our analysis, we consider both sound-wave and turbulence in the case of weak PhT.

\textit{MHD turbulence.--}Usually, the turbulence arises as MHD eddies due to the bubble motion during first-order PhT or some other mechanisms (\emph{e.g.}, \cite{Kamada:2019uxp}) unrelated to PhT. 
The GWB is sourced by the magnetic anisotropy\footnote{From the equation of motion of the metric-tensor perturbation in Fourier space: $\ddot{h}_k + 3 H \dot{h}_k + \frac{k^2}{a^2}h_k = 16\pi G \Pi^{\rm TT}_k$, the energy density of GW is approximate $\rho_{\rm GW} \sim \dot{h}^2/(32\pi G) \sim 8 \pi G \mathcal{T}^2 (\Pi^{\rm TT})^2$ with $\mathcal{T}$ the duration of the source. Neglecting the contribution from fluid motion, the anisotropic stress is dominated by the magnetic field: $\Pi^{\rm TT} \simeq B^2/2$. So the GW energy density at production time is $\rho_{\rm GW} \simeq 2 \pi G B^4 \mathcal{T}^2$.}
and has the energy density fraction at the production: $\Omega_{\rm GW}^* \sim 3 (\Omega_B^*)^2 (H_* \mathcal{T})^2$ \cite{Neronov:2020qrl} where $\Omega_B^* = B_*^2/(2 \rho_{c,*})$ is the magnetic energy-density fraction,  $*$ denotes the time of GW emission, and the sourcing-time scale related to size and velocity of turbulence $\mathcal{T} \sim l_*/v_A \lesssim H_*^{-1}$ with $v_A \simeq \sqrt{2 \Omega_B^*}$ the Alfv{\'e}n speed. The GW energy density becomes $\Omega_{\rm GW}^* \sim \frac{3}{2} \Omega_B^* (H_* l_*)^2$. Nonetheless, depending on the detail of turbulence, the scaling can vary $\Omega_{\rm GW}^* \sim  (\Omega_B^*)^n (H_* l_*)^2$ with $n \in [1,2]$.
The GW spectrum today peaks at amplitude $\Omega_{\rm GW}^{(0)} = \Omega_{\rm rad}^{(0)} \mathcal{G}(T_*) (\Omega_B^*)^n (H_* l_*)^2$ and frequency today $f_{*} = \frac{2}{l_*}\cdot \frac{a_*}{a_0}$ \cite{Neronov:2020qrl}. For simplicity, we consider the case of the largest processed eddies (LPE) \emph{i.e.,} maximized $l_* = v_A /H_*$ and the GW spectrum. The peak position -- the amplitude and frequency -- is related to the magnetic energy fraction and the temperature of the GW production,
\begin{align}
    \Omega_B^{*} &\simeq \left[1.4 \cdot 10^{-5} \left(\frac{h^2\Omega_{\rm GW}^{(0)}(f_*)}{10^{-9}}\right)\left(\frac{10.75}{g_*(T_*)}\right)\left(\frac{g_{*,s}(T_*)}{10.75}\right)^{\frac{4}{3}}\right]^{1/(n+1)},\\
    f_* &\simeq 16.1 ~{\rm nHz} \left(\frac{T_*}{\rm MeV}\right) \left(\frac{10^{-2}}{\Omega_B^{*}}\right)^{1 \over 2},
\end{align}
We can also characterize the MHD turbulence by the magnetic field strength $B_0$ and the magnetic-field correlation length scale $l_B \equiv l_*(a_0/a_*)$ as observed today\footnote{The energy density in magnetic field red-shifts as radiation $\Omega_{B,0} = \Omega_B^* \Omega_{\rm rad} \mathcal{G}(T_*)$, so approximately $B \propto a^{-2}$.},
\begin{align}
    B_0 = 0.4 ~ {\rm \mu G} \left(\frac{\Omega_B^{*}}{10^{-2}}\right)^{\frac{1}{2}} \left(\frac{g_*(T_*)}{10.75}\right)^{\frac{1}{2}}\left(\frac{10.75}{g_{*,s}(T_*)}\right)^{\frac{2}{3}}, ~ ~ l_B \simeq 1.95 ~ {\rm pc} \left(\frac{10 ~ {\rm nHz}}{f_*}\right).
\end{align}

{\bf Cosmic Strings.--} We assume that a string network loses energy only through GW emission of loops. We sum up to $j = 10^5$ modes of loop oscillations. More complicated CS scenarios, \emph{e.g.},~small-loop population~\cite{Lorenz:2010sm} or meta-stable strings~\cite{Buchmuller:2019gfy,Buchmuller:2021mbb,Buchmuller:2020lbh}, can alter the fittings to PTA data, see Ref.~\cite{EPTA:2023hof, NANOGrav:2023hvm}.
We also note that we consider field theory CS from a gauge symmetry, discarding the global symmetry case, whose Goldstone-boson emission leads to a strong $\Delta N_{\rm eff}$ bound~\cite{Chang:2021afa,Gorghetto:2021fsn}.
See \cite{Servant:2023mwt} for the best-fitted result and constraints on global-(axionic)-string SGWB obtained from the recent analysis of the {\tt NG15} data set.

\section{Constraints on scalar-induced gravitational waves}
Large scalar perturbations at small scales, unbounded by CMB, are constrained by their direct consequences and many observations.
We classify two types of constraints:

\textbf{Primordial black holes.}-- 
Sourced by the scalar perturbation, the energy density perturbations collapse into the \emph{primordial black holes} (PBH) \cite{Hawking:1971ei, Carr:1974nx, Carr:1975qj} whose abundance is subjected to the PBH over-abundance and many observational bounds (\emph{e.g.}, gravitational lensing and LVK merger events) \cite{Green:2020jor, bradley_j_kavanagh_2019_3538999}.
 Starting from the scalar perturbation in Eq.~\eqref{eq:gaussian_scalar}, we use the \emph{peak theory} method in  Ref.~\cite{Young:2019yug} with a \emph{real-space top-hat} window function (with $\gamma = 0.36$, $\kappa = 4$, and $\delta_c = 0.55$) to calculate the total PBH abundance today, parametrized by $f_{\rm PBH}^{\rm tot}$ the total fraction of the \emph{cold dark matter} today.
 We also include the non-linear effect between the primordial scalar and the energy density perturbations.
 
 Since the collapse happens when the scalar perturbation re-enters the horizon, many populations of PBH can be formed from a broad scalar power spectrum along the cosmic history. 
 However, the spectral shape of the PBH abundance over mass range inherits that of the primordial scalar perturbation. For the SIGW peaked at frequency $f_*$, the scale of the scalar-perturbation peak re-enters the horizon and collapses into the peak of PBH mass distribution with mass defined by the horizon mass $M_H$, \emph{i.e.,} the total energy density  enclosed within the Hubble volume at temperature $T$,
 \begin{align}
     M_H =  \frac{4\pi}{3}H^{-3} \left(3 \Mpl^2 H^2 \right) \simeq 44 \, M_{\odot} \, \left(\frac{g_{*}(T)}{10.75}\right)^\frac{1}{2} \left(\frac{10.75}{g_{*s}(T)}\right)^\frac{2}{3}\left(\frac{10^{-9} \,{\rm Hz}}{f_*}\right)^2,
     \label{eq:horizon_mass}
 \end{align}
 where $M_\odot \simeq 1.988 \times 10^{30}$ kg is the Solar mass, and we used the frequency-temperature relation \eqref{eq:freq_temp_relation}.
 For a typical frequency of SIGW in PTA ($f_* \simeq 1-10$ nHz), the PBH is populated in the mass range of $M_{H} \simeq \mathcal{O}(0.4-40) M_\odot$.
 
\begin{figure*}[h]
    \centering
    {\bf 2nd Order Scalar Induced GWB (Gaussian)}\\
    \includegraphics[width=\linewidth]{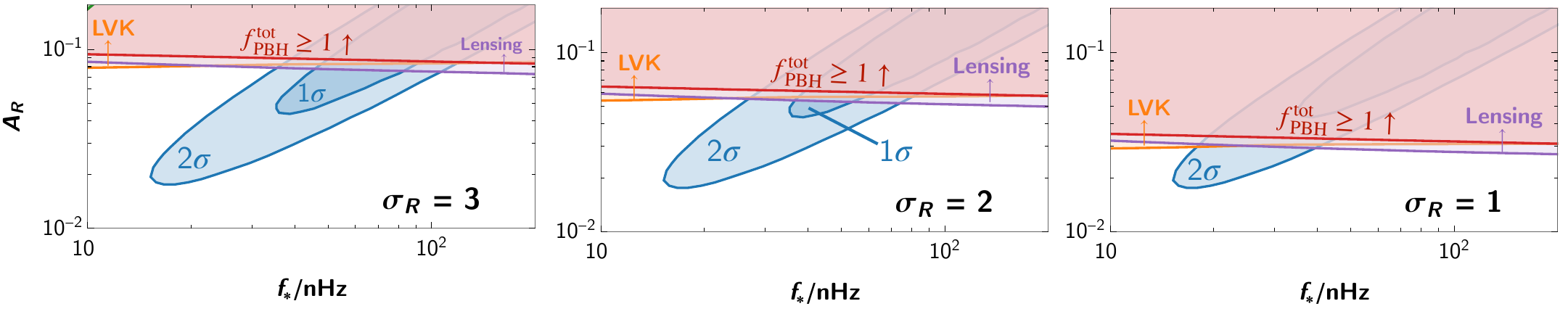}
    \caption{Constraints on the Gaussian scalar-perturbation bumps in Eq.~\eqref{eq:gaussian_scalar} from PBH abundance (over-abundance, lensing, LVK mergers) and other constraints ($\mu$-distortion). The parameters are $\lbrace \sigma$, $f_{*}$, $A_{\mathcal{R}}\rbrace$.
    The 2OSI GW interpretation of the PTA signal (in blue) survives but cannot account for the dark matter totality.}
    \label{fig:SIGW_constraint}
\end{figure*}

 The most conservative constraint is the PBH over-abundance $f_{\rm PBH}^{\rm tot} > 1$, 
 while many observations also exclude the PBH abundance for $f_{\rm PBH}^{\rm tot} < 1$ over a wide range of masses \cite{Green:2020jor, bradley_j_kavanagh_2019_3538999}.
 There are two dominant constraints for the Solar-mass PBH compatible with PTA interpretation: the LVK mergers, individual events and stochastic background, and the gravitational lensing constraints.
 We translate the bounds on the monochromatic PBH mass in Ref.~\cite{Green:2020jor} using the method outlined in Ref.~\cite{Carr:2017jsz} for our extended PBH mass distribution. 
 Fig.~\ref{fig:SIGW_constraint} shows the constraints (PBH over-abundance, lensing, and LVK mergers) on the log-normal scalar perturbation.
 We also calculate the PBH abundance in the presence of the pNG scalar perturbation by applying the simplifying technique of Ref.~\cite{Young:2022phe} to the procedure of the Gaussian case. 

\textbf{CMB and $\mu$-distortion.}-- 
Though the scalar perturbation power spectrum peaks at the scale of PTA, its tail at scales larger than PTA ($k \lesssim 10^6 ~ {\rm Mpc}^{-1}$) can be constrained for a too broad spectrum. First, the sizeable scalar perturbation sources the density fluctuation, dissipating into photon fluid and causing the $\mu$-distortion of the CMB spectrum \cite{Chluba:2012we,Kohri:2014lza}. Another constraint arises when the spectrum is broad and large to contradict the CMB observations \cite{Hunt:2015iua}.
By translating the bounds in Ref.~\cite{Inomata:2018epa}, we find no such bound entering the shown parameter spaces of Fig.~\ref{fig:SIGW_constraint}.

\begin{table*}[h!]
	\begin{tabular}{ccl}
		\hline\\ [-0.75em]
	 \textbf{Parameter} 	&  \textbf{Description}  &  \textbf{Prior}  \\ [0.25em] \hline\\[-0.5em]
        & \textbf{Power law (PL)} & \\[0.5em]
		$\mathcal{A}^{(*)}_{\rm inf}$ &  Amplitude of the signal at $f_* = f_{1 \rm yr}$ & log-uniform [$10^{-20}, 10^{-4}$]  \\
		$n_t$ &  Spectral tilt  & uniform [$-4, 6$]  \\[0.5em] \hline \\[-0.5em]
	   & \textbf{SMBHBs (PL $n_t = 2/3$)} &  \\ [0.5em]
		$\mathcal{A}^{(*)}_{\rm 2/3}$ & \hspace{.5cm} Amplitude of the signal at $f_* = f_{1 \rm yr}$ \hspace{.5cm} & log-uniform [$10^{-14}, 10^{-6}$] \\
		[0.5em] \hline \\ [-0.5em]
        & \textbf{Flat (PL $n_t = 0$)} &  \\ [0.5em]
		$\mathcal{A}^{(*)}_{\rm 0}$ & \hspace{.5cm} Amplitude of the signal at $f_* = f_{1 \rm yr}$ \hspace{.5cm} & log-uniform [$10^{-13}, 10^{-4}$] \\
		[0.5em] \hline \\ [-0.5em]
        & \textbf{Field-theory Cosmic strings (field th. CS)} & \\[0.5em]
		$G\mu$ &  Cosmic-string tension & log-uniform [$10^{-14}, 10^{-6}$]  \\[0.5em] \hline \\[-0.5em]
        & \textbf{Cosmic super-strings (super CS)} & \\[0.5em]
		$G\mu$ &  Cosmic-string tension & log-uniform [$10^{-14}, 10^{-6}$]  \\
        $p$ &  Inter-commutation probability & log-uniform [$10^{-3}, 1$] 
        \\[0.5em] \hline \\[-0.5em]
        & \textbf{Broken power law (BPL)} & \\
		$\mathcal{A}_*$ & Peak amplitude & log-uniform [$10^{-14},10^{-7}$] \\
        $f_*$ [nHz] & Peak frequency & log-uniform [$10^{-2.5}, 10^{4.5}$] \\
        $n_1$ & Low-frequency tilt $f\ll f_*$ & uniform [$-8,8$] \\
        $n_2$ & High-frequency tilt $f\gg f_*$ & uniform [$-8,8$] \\[0.5em] \hline \\[-0.5em]
        & \textbf{Audible axion (AA)} & \\[0.5em]
		$f_a$ [$m_p$]& Axion decay constant  & log-uniform [$10^{-4}, 10^2$]  \\
		$m_a$ [{\rm meV}]&  Axion mass  & log-uniform [$10^{-13},10^{-9}$] \\[0.5em] \hline \\[-0.5em]
        & \textbf{Gaussian 2$^{\rm nd}$ order scalar-induced GWB: bump (gSIGWB)} & \\[0.5em]
		$\mathcal{A}_\mathcal{R}$ & Amplitude of the scalar power spectrum  & log-uniform [$10^{-3}, 10$]  \\
		$\sigma_\mathcal{R}$ &  Variance of the scalar power spectrum  & log-uniform [$0.1, 3$]  \\
        $f_*$ [nHz] &  Frequency corresponding to $k_*$  & log-uniform [$10^{-2}, 10^4$]  \\[0.5em] \hline \\[-0.5em]
        & \textbf{Non-gaussian 2$^{\rm nd}$ order scalar-induced GWB: bump (ngSIGWB)} & \\[0.5em]
		$\mathcal{A}_\mathcal{R}$ & Amplitude of the scalar power spectrum  & log-uniform [$10^{-3},10$]  \\
		$\sigma_\mathcal{R}$ &  Variance of the scalar power spectrum  & log-uniform [$0.025,0.2$]  \\
        $f_*$ [nHz] &  Frequency corresponding to $k_*$  & log-uniform [$10^{-1}, 10$]  \\
        $f_\texttt{nl}$ &  Local non-Gaussianity (and imposing the perturbativity: $A_\mathcal{R}f_{\tt nl}^2 \lesssim 1$ \cite{Adshead:2021hnm})  & log-uniform [$10^{-5},10^5$]  \\[0.5em] \hline \\[-0.5em]
        & \textbf{Strong first-order PhT: bubbles ($\alpha \gg 1$ runaway in vacuum) (PhTRV)} & \\[0.5em]
		$\beta/H_*$ & Inverse transition duration  & log-uniform [$1,500$]  \\
		$T_*$ [GeV] &  Transition temperature  & log-uniform [$10^{-3}, 10^5$]  \\[0.5em] \hline \\[-0.5em]
        & \textbf{Weak first-order PhT: sound-wave and turbulence (non-runaway) (PhTNR)} & \\[0.5em]
        $\alpha$ & Transition strength  & log-uniform [$10^{-2},1$]  \\
		$\beta/H_*$ & Inverse transition duration  & log-uniform [$1,500$]  \\
		$T_*$ [GeV] &  Transition temperature  & log-uniform [$10^{-3},10^5$]  \\[0.5em] \hline \\[-0.5em]
        & \textbf{Magnetro-hydrodynamics (MHD) turbulence} & \\[0.5em]
		$T_*$ [GeV] & Temperature when turbulence is generated  & log-uniform  [$10^{-3},10^6$]  \\
        $B_0$ [$\mu$G] & Magnetic field strength  & log-uniform  [$10^{-3},10^2$]  \\
		$l_B$ [pc] &  Magnetic field correlation length  & log-uniform  [$10^{-6},10^3$]  \\[0.5em] \hline
	\end{tabular}
	\caption{Summary of all priors}
	\label{tab:priors}
\end{table*} 

\newpage

\end{document}